\documentclass{ws-ijmpd}
\usepackage[super,compress]{cite}

\def\rs{\rm s}
\def\rs1{\rm s$^{-1}$}

\def\rcm{\rm cm}
\def\rcm2{\rm cm^{-2}}

\def\c2r{\chi^2_\nu}
\def\chisq{\chi^2}
\def\eze{$E_{\rm 0}$ } \def\eze{$E_{\rm 0}$}
\def\epo{$E_{\rm p}$}
\def\epi{\ensuremath{E_{\rm p,i}}}
\def\eiso{\ensuremath{E_{\rm iso}}}
\def\ega{\ensuremath{E_{\rm jet}}}
\def\eiso{\ensuremath{E_{\rm iso}}}
\def\liso{\ensuremath{L_{\rm iso}}}
\def\lpiso{\ensuremath{L_{\rm p,iso}}}
\def\ega{\ensuremath{E_{\gamma}}}
\def\epeiso{$E_{\rm p,i}$ -- $E_{\rm iso}$}

\def\nufnu{$\nu F_{\nu}$ }
\def\epega{$E_{\rm p,i}$ -- $E_{\gamma}$ }

\def\sext{$\sigma$$_{\rm ext}$}
\def\sax{{\it Beppo}SAX }
\def\swift{{\it Swift}}
\def\fermi{{\it Fermi}}

\def\omegam{$\Omega$$_{\rm M}$}
\def\wzero{$w_{\rm 0}$}
\def\wa{$w_{\rm a}$}
\def\omegal{$\Omega$$_{\Lambda}$}
\def\h0{H$_{\rm 0}$~}
\def\tb{t$_{\rm b}$~}
\begin{document}

\markboth{Amati \& Della Valle}
{Measuring cosmological parameters with GRBs}

%
\catchline{}{}{}{}{}
%

\title{MEASURING COSMOLOGICAL PARAMETERS WITH GAMMA RAY BURSTS\footnote{Based on a talk presented at the Thirteenth Marcel Grossmann Meeting on General Relativity, Stockholm, July 2012.}}

\author{LORENZO AMATI}

\address{INAF - IASF Bologna, via P. Gobetti 101\\ 40129 Bologna, Italy\\
amati@iasfbo.inaf.it}

\author{MASSIMO DELLA VALLE}

\address{INAF - Osservatorio di Capodimonte, Salita Moiariello 16 \\ 80131 Napoli, Italy \\
dellavalle@na.astro.it}

\maketitle

\begin{history}\received{8 August 2013} \revised{10 August 2013}
\end{history}

\begin{abstract}

 In a few dozen seconds gamma ray bursts (GRBs) emit up to 
$\sim$10$^{54}$ erg in terms of an equivalent isotropically radiated energy \eiso,
so they can be observed  up to $z\sim 10$. Thus, these phenomena appear to be very
promising  tools  to describe the expansion rate history  of the universe. 
Here we review the use of the \epeiso{} correlation
of GRBs to measure 
the cosmological density parameter $\Omega_M$.
We show that the present data set of gamma ray bursts, coupled with the
assumption that we live in a flat universe, can provide independent
evidence, from other probes, that \omegam{}$\sim$0.3.  We show that current 
(e.g., \swift,  \fermi/GBM, Konus-WIND) and
forthcoming GRB experiments 
(e.g., CALET/GBM, SVOM, Lomonosov/UFFO, LOFT/WFM) will allow us to constrain
\omegam{} with an accuracy
comparable to that currently exhibited by Type Ia supernovae and to study the properties of dark energy
and their evolution with time. 
\end{abstract}

\keywords{cosmological parameters, gamma-rays: bursts, gamma rays: observations}
\ccode{PACS numbers:95.36.+x;95.85.Pw;98.70.Rz;98.80.Es}

\section{Introduction}	

The Hubble diagram of Type Ia supernovae (SNe--Ia) observed at the end of 90's
was hard to reconcile with the decelerated trend implied by the 
Einstein-de Sitter model, then suggesting an accelerated expansion of
the universe\cite{Perlmutter98, Perlmutter99,Riess98, Schmidt98}. In the following decade, both  SNe--Ia
\cite{Tonry03,Knop03,Astier06,WW07,Kowalski08} 
and other cosmological probes, such as the cosmic microwave background 
(CMB)\cite{Debernardis00,Spergel03,Dunkley09,Komatsu11,Ade13}, 
galaxy clusters and baryonic acoustic oscillations (BAO)\cite{Eisenstein05,Percival10,Suzuki12} lent further support to the 
existence of an unknown form of ``dark energy'' propelling the acceleration. The 
equation of state of such dark energy is often expressed as $P=w(z)\rho$, where 
$w(z)= w_{0} + {w_{a}z/( 1+z)}$.\cite{Chevalier01,Linder03}

After combining SNe data with flat universe constraints from CMB measurements 
\cite{Debernardis00,Spergel03}, Riess et al. (2004)\cite{Riess04}
found $w_0$ $\sim -1$ and $w_a$ $\sim 0$. 
These results identified 
the ``dark energy'' with the cosmological constant. However, one main issue (among 
the others) remains still unsolved: a cosmological constant interpreted as a 
vacuum energy is about 120 orders of magnitude smaller than its
'natural' value computed from quantum mechanics. 
\cite{Carroll06} This fact justifies a lot of interest in models where the 
present energy density (or a dominant fraction of it) is slowly varying 
with time.

In this short review we show that gamma ray bursts (GRBs)
can significantly contribute to shedding some light on this last issue. 
Due to their huge energetic outputs, up to $\sim$10$^{54}$ erg in terms of equivalent  isotropically
radiated energies, released in a few tens or hundreds of seconds 
(Fig.~1), GRBs are the brightest cosmological sources in the universe.
\cite{Piran04,Meszaros06,Gehrels09,Zhang11} 
Therefore,  they have been observed up to $z\sim8-9$ \cite{Salvaterra09,Cucchiara11} (Fig.
1), well beyond the observing redshift range of SNe--Ia, limited to $z<2$. 
\cite{Riess07,Suzuki12,Jones13} In addition, GRBs emit most of their 
radiation as hard X-rays, thus they are only marginally affected by
uncertainties connected with correction for reddening.\cite{Dellavalle92,Folatelli10}
In recent years it has been 
shown \cite{Amati08,Dellavalle09} that the robust correlation between 
the photon energy at which the \nufnu{} spectrum peaks and the GRB radiated energy 
\cite{Amati02,Amati06} can be successfully 
used to measure 
the cosmological density parameter \omegam.

\begin{figure}[t]
\centerline{\includegraphics[width=0.59\columnwidth]{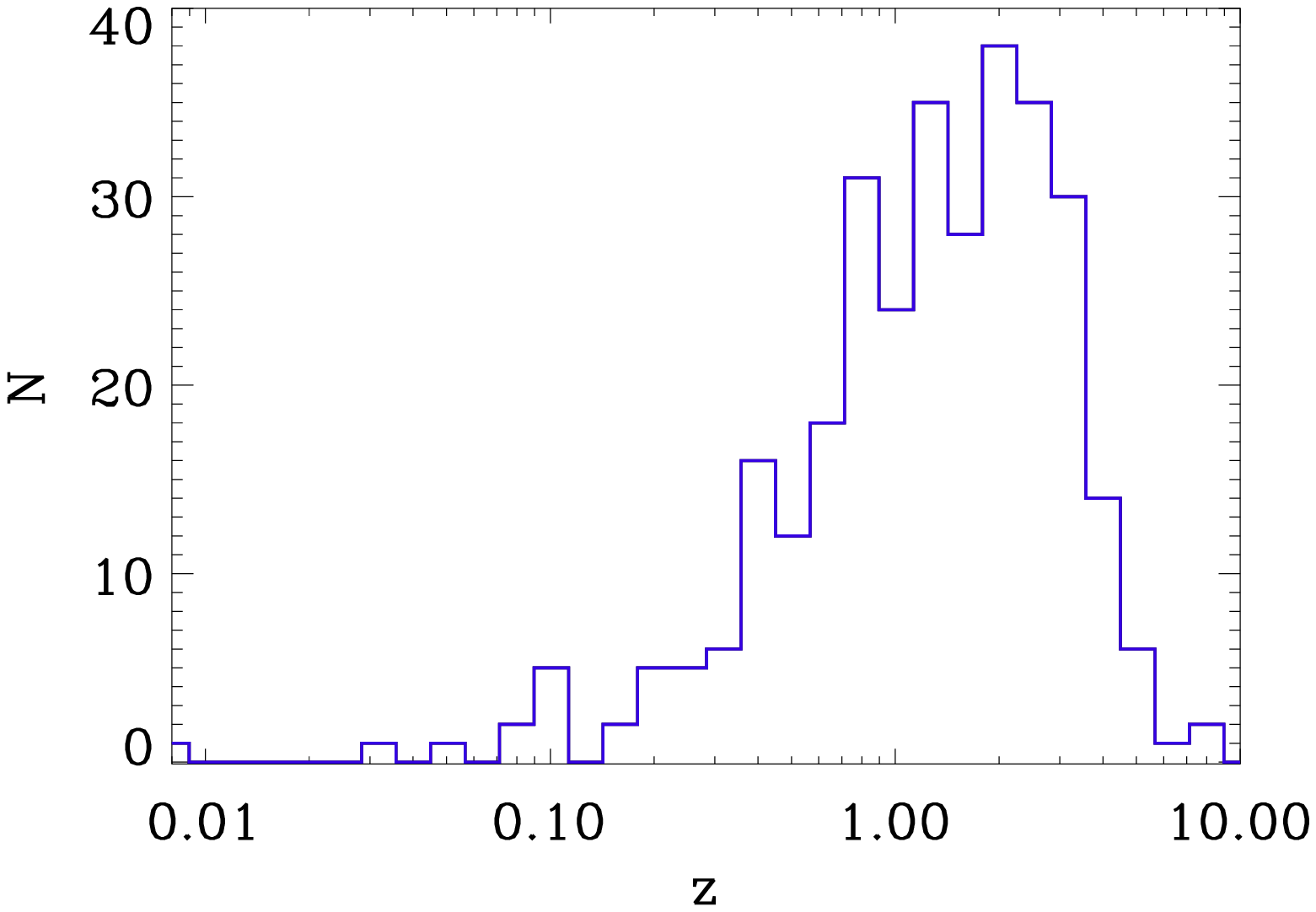}
\hspace{-0.6cm}
\includegraphics[width=0.59\columnwidth]{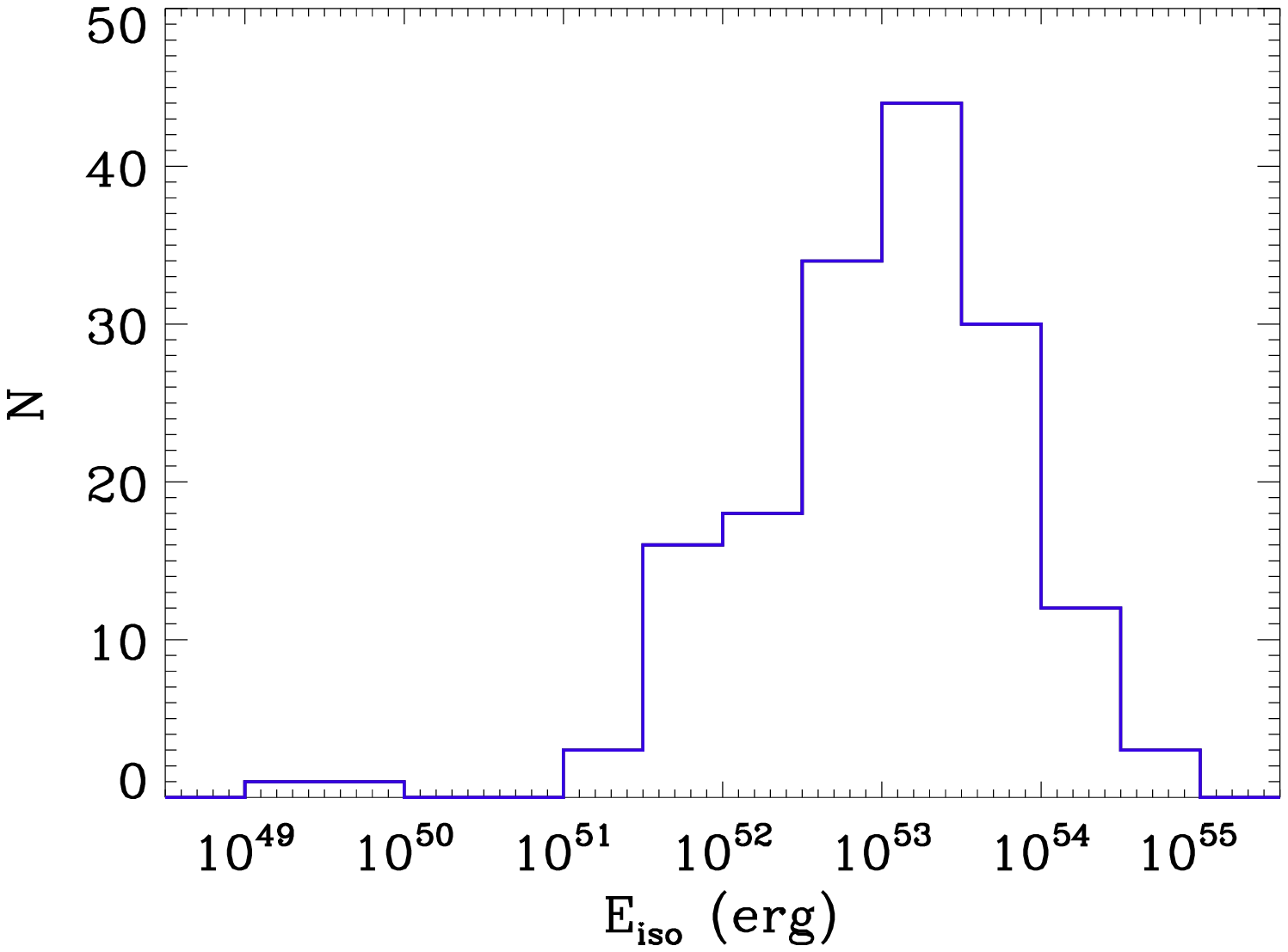}}
\caption{Distributions of redshift (left panel) and of equivalent isotropically 
radiated energy \eiso 
(computed by assuming
the standard flat FLRW cosmology with \h0=70 and \omegam=0.3) of GRBs as of
end 2012.
}
\end{figure}

\begin{figure}[t]
\centerline{\includegraphics[width=0.59\columnwidth]{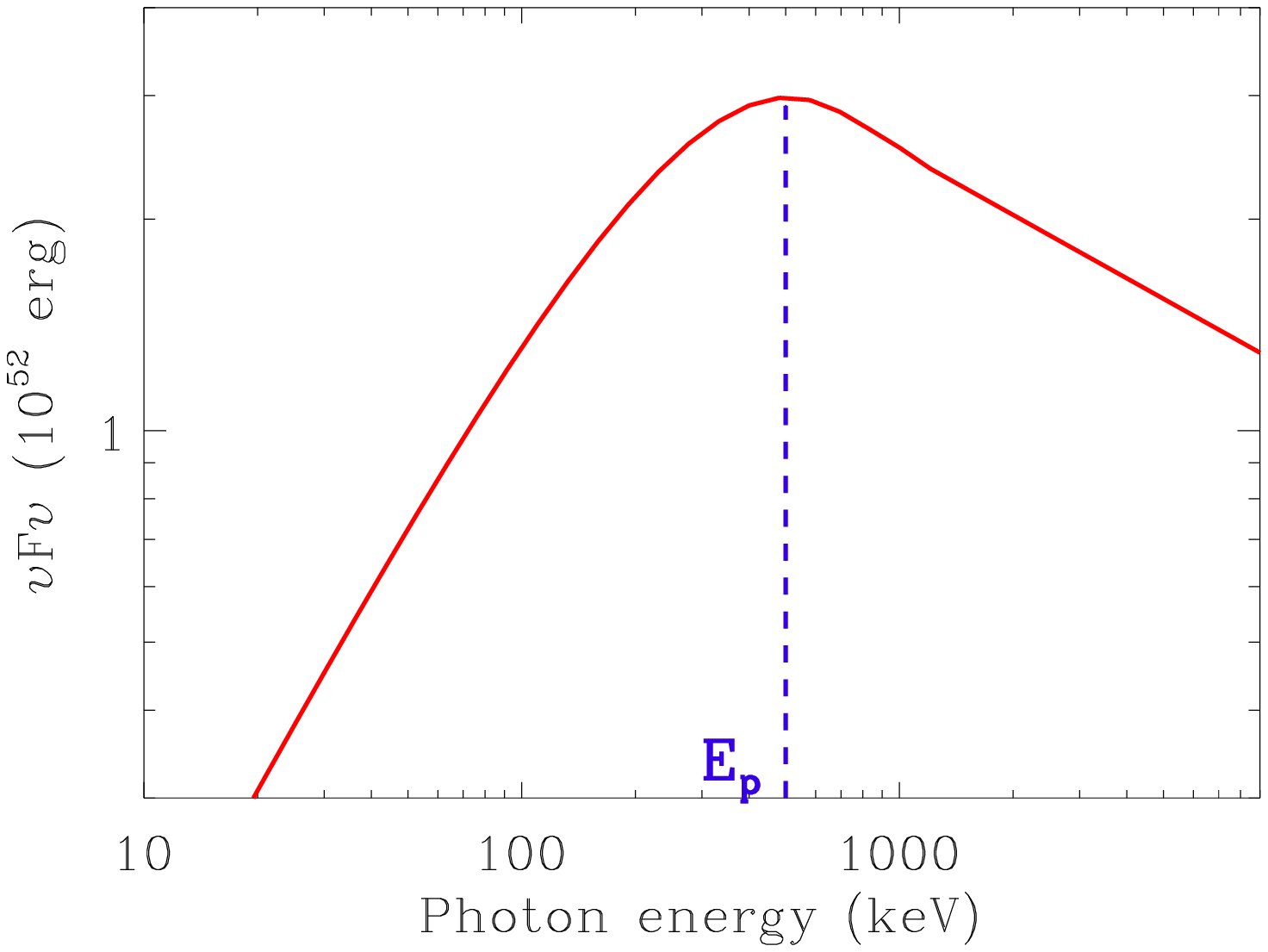}
\includegraphics[width=0.59\columnwidth]{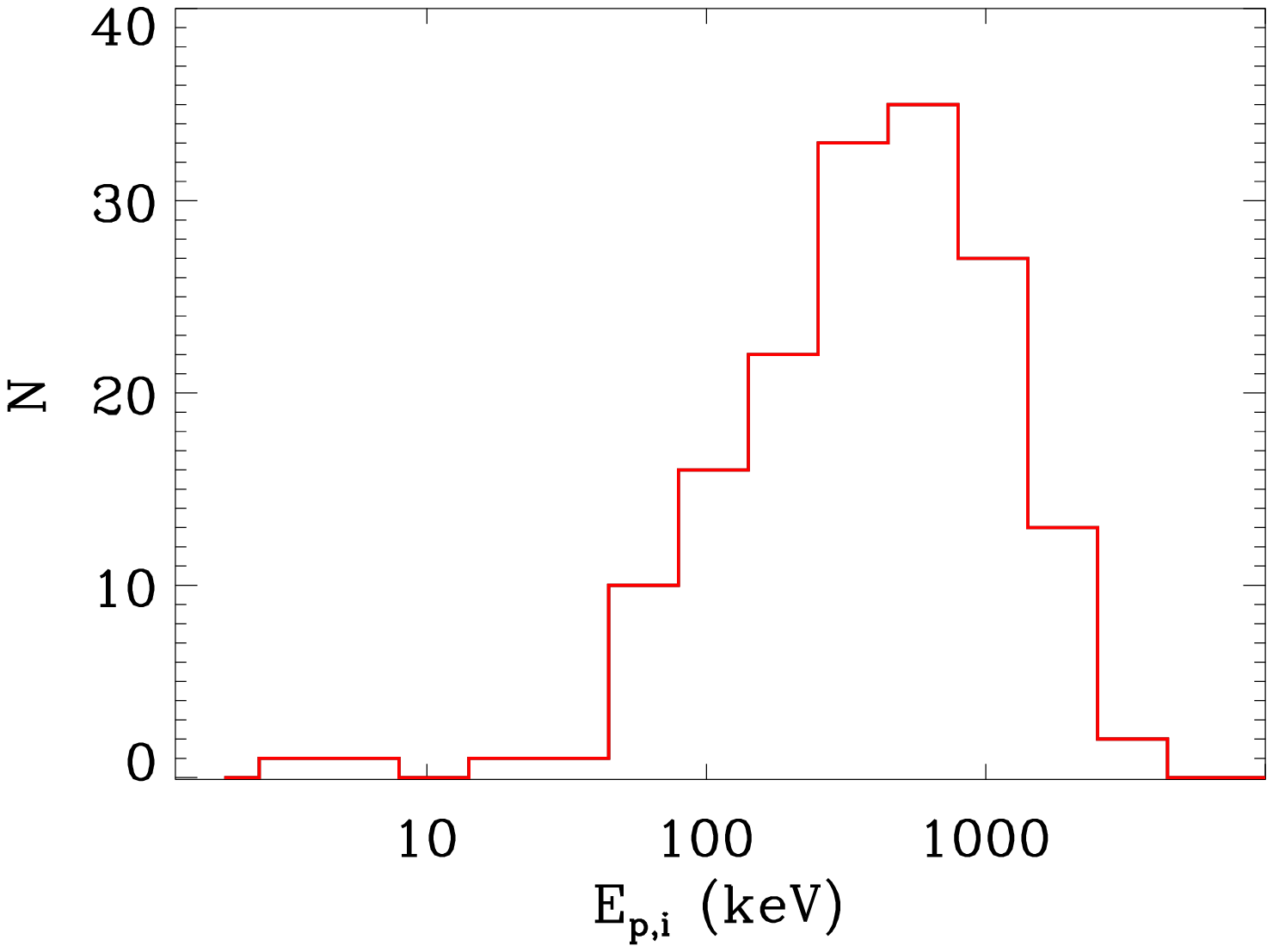}}
\caption{Left: typical \nufnu (energy) spectrum of a GRB in its cosmological 
rest frame. Right: distribution of the photon energy \epi\ at which the cosmological
rest frame \nufnu spectrum of GRB reaches its maximum.
}
\end{figure}

\section{Spectra and energetics of GRBs}

The spectrum of GRBs is nonthermal and can be empirically described by the 
so-called Band function, i.e., a smoothly broken power law
characterized by three parameters: the
low-energy spectral index $\alpha$, the high energy spectral index $\beta$ and the 
"rollover" energy \eze{}.\cite{Band93} As can be seen in Fig.~2,
the \nufnu{} spectrum shows a peak; the photon energy at which this peak occurs is a 
characteristic quantity in GRB emission models and is called the ``peak energy" \epo. 

Every GRB for which it is possible to measure the redshift and the spectrum can be characterized by 
two key parameters: the total radiated energy, computed by integrating the spectrum in a 
standard 1--10000 keV energy band and assuming isotropic emission, \eiso, and the 
peak energy of the cosmological rest-frame \nufnu spectra of GRBs: {\rm \epi{} = 
\epo{} $\times$~$(1 + z)$}. The updated distribution (as of the end of 2012) of
\epi{} for 156 events  (Fig.~2) is approximately a Gaussian centered at a few hundreds of keV and 
with a low energy extension down to a few keV, corresponding to the so-called 
X-ray flashes (XRFs) or X-ray rich (XRR) events. The distribution of \eiso{} is 
somewhat similar, extending from $\sim$10$^{48}$ to more than $\sim$10$^{54}$ erg, 
and peaking at $\sim$10$^{52}$ erg (Fig.~1).

The distribution of \eiso{} as a function of redshift (Fig.~3) 
shows the lack of detection of weak events at high redshift (which is expected as 
result of an obvious bias due to instruments detection threshold) while the lack of bright events at 
low redshifts may be the result of two factors: an intrinsically low rate 
of events (less than 3\% of  SNe--Ibc are associated with long duration 
GRBs \cite{Guetta07}) combined with a correlation between the jet opening angle and GRB 
intensity (the smaller the
opening angle is, the brighter the GRB appears to be). \cite{Amati07,Ghirlanda13}  

\begin{figure}[t]
\centerline{\includegraphics[width=0.65\columnwidth]{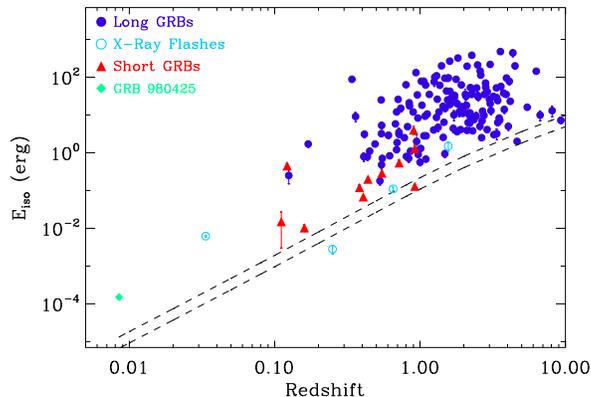}}
\caption{Distribution of isotropic-equivalent 
radiated energy, \eiso, as a function of redshift
for different classes of GRBs (as of end 2012). The dashed lines in the right
panel show typical instrumental sensitivity limits as a function of redshift
(updated from Amati \& Della Valle 2013\cite{Amati13a}). 
}
\end{figure}

\section{The \epeiso{} correlation}
 
In 2002, based on a small sample of \sax{} GRBs with known redshift and spectral 
parameters, it was discovered that \epi{} is significantly 
correlated with \eiso{}.\cite{Amati02,Amati06} This correlation 
has the form 
$$
\log{\epi({\rm keV)}} = m\log{\eiso(10^{52}{\rm 
erg})} + q \,,
$$
with $m$$\sim$0.5 and $q$$\sim$2, and is characterized by an 
extra-Poissonian scatter normally distributed with a \sext{} of $\sim$0.2 dex 
around the best fit law. Subsequent observations 
with various detectors and spectrometers confirmed and extended the \epeiso{}
correlation (Fig.~4), showing that it holds for all long GRBs and XRFs with well
measured redshift and spectral parameters.\cite{Amati06,Amati08}
The data, together 
with the power law best fitting the long GRB points and their confidence regions, 
have been taken from Amati et al.~(2013).
\cite{Amati13b} 

As discussed by several authors, the existence of the \epeiso{} relation, its
extension over several
orders of magnitude in both coordinates, its slope and its intrinsic dispersion
are a challenging observational evidence for the current
models of the physics and geometry of the prompt  emission of GRBs.\cite
{Zhang02,Lamb04,Amati06} 
In addition, the \epeiso{} plane can be used to identify different classes of GRBs (short
vs.\ long)  and get clues on 
their different nature. For instance, 
the consistency of XRFs with the \epeiso{} correlation 
strongly supports the idea that they are not a different class of sources but 
form the faint tail of the ``cosmological GRB'' population. There is also clear 
evidence that short GRBs, for which redshift estimates became available only in the 
last few years, do not follow the correlation, indicating that their main emission 
mechanism, and/or the geometry of their emission, may be different from long ones. 
Finally, the weak long GRB\,980425 (only 
40 Mpc away)  which is also the ``prototype" event for the GRB--SN connection, is 
inconsistent with the \epeiso{} correlation, suggesting the possible existence of a 
``local" population of sub-energetic GRBs with different properties with respect to 
cosmological long GRBs.\cite{Soderberg06,Amati07,Guetta07}

\begin{figure}[t]
\centerline{\includegraphics[width=0.75\columnwidth]{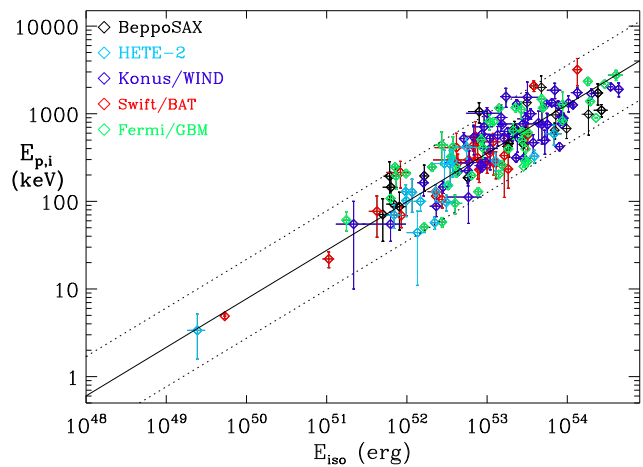}
}
\caption{The \epeiso{} correlation in long GRBs (as of the end of 2012). The black line shows the best fit power law. For each point, the color identifies the instrument which
performed the spectral measurement.
}
\end{figure}

\section{Reliability of the \epeiso{} correlation}

Different GRB detectors are characterized by different thresholds and spectroscopic 
sensitivity, therefore they can spread relevant selection effects / biases in the 
observed \epeiso{} correlation. In the past, there were 
claims that a high fraction (70--90\%) of BATSE GRBs without redshift would be 
inconsistent with the correlation for any redshift.\cite{Band05,Nakar05} 
However, this ``peculiar'' conclusion was refuted by other authors 
\cite{Ghirlanda05,Bosnjak08,Ghirlanda08,Nava11} who show that, in
fact,  most BATSE GRBs with 
unknown redshift were well consistent with the \epeiso{}
correlation. We also note 
that the inconsistency of  such a high percentage of GRBs of unknown redshift would 
have implied that most GRBs with known redshift should also be inconsistent with the 
\epeiso{} relation, and this fact was never observed. Moreover, Amati et al. (2009)\cite{Amati09} showed 
that the normalization of the correlation varies only marginally using GRBs 
measured by individual instruments with different sensitivities and energy bands, 
while Ghirlanda et al. (2010) \cite{Ghirlanda10} show that the parameters of the 
correlations ($m$ and $q$) are independent of redshift.

Furthermore, the \swift{} satellite, thanks to its capability of providing quick and accurate localization of 
GRBs, thus reducing the selection effects in the observational chain leading to
the estimate of GRB
redshift, has further confirmed the reliability of the \epeiso{}
correlation.\cite{Amati09,Ghirlanda10,Sakamoto11}

Finally, based on time-resolved analysis of BATSE, \sax and \fermi{} GRBs, it was 
found that the \epeiso{} correlation also holds within
each single GRB with
normalization and slope consistent with those obtained with time-averaged spectra
and energetics / luminosity.
\cite{Ghirlanda10,Lu12,Frontera12,Basak13} This ultimate test
confirms the physical origin 
of the correlation, also providing clues to its explanation.

\subsection{K Correction}

The instruments that measure GRB fluences cover an energy range from $\sim$10 keV 
to $\sim$10 MeV. The spectral energy distribution of GRBs peaks typically 
at hundreds keV, therefore the effect of using the finite spectral band of the 
detectors on measuring \epi{} and  \eiso{} is negligible. We have estimated 
\eiso{} to be affected by this bias by no more than 10\%.\cite{Amati09} 
This conclusion is also confirmed by the following facts.
\begin{itemize}
\item Detectors with different energy bands and sensitivity provide the same \epeiso{}
correlation (see Section 4 and Fig.~4). In addition, cross--calibration of the 
instruments is taken into account and included in the uncertainties on \epi{} and
\eiso.
\item{} By dividing the GRB sample into subsets with different redshift ranges (e.g., 
$0.1 < z < 1$, $1 < z < 2$, etc.), it is found that slope, normalization and dispersion 
of the correlation do not change significantly along z.\cite{Ghirlanda08} This result 
also implies that Malmquist--like selection effects are negligible.
\end{itemize}

\section{The beginning of GRB cosmology}

The idea to use GRBs as cosmological rulers was proposed in 2004, when it was 
found that the \epeiso{} correlation tightened when \eiso{} 
was replaced with the collimation-corrected radiated energy $\ega{} = (1 - 
\cos{\theta_{jet}}) \times \eiso$.\cite{Ghirlanda04,Dai04}
This result was based on a small subsample of  GRBs with known
\epi{} and \eiso{}  for which it was possible to infer the jet 
opening angle $\theta_{jet}$ from the ``break time" \tb\ at which the decay  
of the light curve of the optical afterglow becomes steeper.
\cite{Sari99} By exploiting  the low scatter of the \epega{} correlation and applying statistical methods 
accounting for the lack of calibration, it was 
possible to derive, within the ``standard" FLRW cosmological model, 
estimates of \omegam{} and \omegal{} consistent with the ``concordance" values 
mostly coming from the analysis of Type Ia SNe and the CMB. A review of these 
methods and results is provided in Ghirlanda et al. (2006).\cite{Ghirlanda06} The 
results of these investigations were encouraging, and prompted other studies aimed 
at deriving a GRB Hubble diagram, based, e.g., on the joint use of the \epega{} 
correlation together with other weaker correlations between luminosity and observed 
properties \cite{Schaefer07}, or the calibration of the \epega{} correlation with 
Type Ia SNe.\cite{Kodama08,Demianski12}

However, in the last years, the simple jet model assumed to compute $\theta_{jet}$ 
from the break time of the optical afterglow light curve has been questioned
on the basis of different behaviors exhibited by the X-ray afterglow light curves. 
In many cases  achromatic ``jet-breaks" were not detected.\cite
{Panaitescu06,Curran08,Liang08} This fact makes the determination of \ega, and thus the 
characterization and use of the \epega{} correlation, less firm and, in any case,
model dependent.

\begin{figure}[t]
\centerline{
\includegraphics[width=0.75\columnwidth]{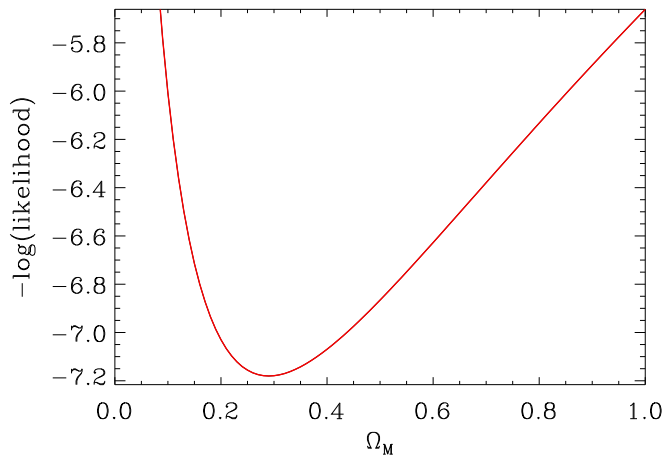}}
\caption{Goodness of fit, in terms of normalized $-log$(likelihood), of the \epeiso{} 
correlation
of long GRBs as a
function of the value of \omegam{} used to compute the \eiso{} values
in a flat FLRW universe (from Amati et al. 2013\cite{Amati13b}). See the text for a description
of the maximum-likelihood method adopted.
}
\end{figure}

\section{The GRB cosmology through the \epeiso{} correlation}

In view of the previous drawbacks,  the possibility to
measure  the cosmological parameters through the \epeiso{} correlation was investigated \cite{Amati08},
which has the advantage,  with respect to the \epega{} correlation, of:
i) allowing for a GRB sample about four times larger because it is based on
only two observables (the  \epega{} correlation requires, in addition to
\eiso{} and \epi, also $t_b$) ii) Unlike of \ega{} computation, the use
of the \epeiso{} correlation it is not model dependent. Indeed no 
assumptions about the jet model, afterglow model, density and
profile of the circumburst environment, or efficiency of conversion of fireball
kinetic energy into radiated energy are needed. 

The question is: how to measure \omegam{} if we need \omegam{} to compute the luminosity distance  and in turn \eiso? 
We can avoid this ``circularity problem''  after floating the value of \omegam\ between 0 and 1 and assuming that the
dispersion of the \epeiso{} correlation will minimize corresponding to
the choice of the ``correct'' value.  

Based on a sample of 70 long GRBs with known \epi{} and \eiso, it was found that, 
after assuming a flat universe, the $\chisq$ obtained by fitting the correlation with 
a simple power law is a function of the value of \omegam{} assumed in the 
computation of \eiso.\cite{Amati08} 
However, given the significant extra-Poissonian scatter of the correlation, the simple 
$\chisq$ method cannot be used to obtain reliable confidence levels on both the 
parameters of the correlation (normalization and slope) and the cosmological 
parameters. In order to get rid of this, Amati et al. (2008)\cite{Amati08} 
adopted a maximum 
likelihood method \cite{Dagostini05,Reichart01}
that takes into account the uncertainties in both the 
$X$ and $Y$ quantities and the extra variance \sext. 
Remarkably, as can be seen in Fig.~5, the 
$-log$(likelihood)
shows a nice parabolic shape, with a minimum at \omegam$\sim$0.30.  This a very 
simple but relevant result: a) it shows that the \epeiso{} correlation can indeed 
be used to extract clues on the values of cosmological parameters; b) it provides evidence, 
independently of Type Ia SNe, that \omegam{} is significantly smaller than 1 and
around 0.3.

In Table~1,  we show the 68\% c.l. intervals for \omegam{} and \wzero{} in a flat FLRW
universe derived with the 70 GRBs of Amati et al. (2008),
the 156 GRBs available as of the end of 2012 (Amati et al. 2013\cite{Amati13b}), a sample
of 250 GRBs (156 real + 94 simulated), expected to be available within a few years, and a sample
of 500 RBs (156 real + 344 simulated) which may be expected from
future dedicated space missions (see next section). 
These values were obtained with the same approach 
as Amati et al. (2008) but using the likelihood function proposed by Reichart (2001)\cite{Reichart01}, which
has the advantage of not requiring the arbitrary choice of an independent
variable among \epi{} and \eiso{}. Interesting enough, we note that, after increasing 
the number of GRBs from 70 to 156, the accuracy of the estimate of \omegam{} improves by a factor of $\sim \sqrt{N_2/N_1}$. The accuracy of these measurements is still lower 
than obtained with supernova data, but promising in view of the increasing number of 
GRBs with measured redshift and spectra (see also Fig.~6, Fig.~7 and the next section). 

\begin{table}[t]
\tbl{Comparison of the 68\% confidence intervals on \omegam{} and \wzero{} 
(\omegam=0.3, \wa=0.5) for a flat FLRW 
universe obtained with the sample of 70 GRBs by Amati et al. (2008), the 
sample of 156 GRBs available as of the end of 2012 (Amati et al.\ 2013\cite{Amati13b}) and
simulated samples of 250 and
500 GRBs (see text).  In the last three lines we also show the results obtained
for the same samples by assuming that the slope and normalization of the \epeiso{} correlation 
are known with a 10\% accuracy based, e.g., on calibration against SNe--Ia or self-calibration
with a large enough number of GRBs at similar redshift.}
{\centerline{
\begin{tabular}{lcc} 
\hline
 GRB \# & \omegam{} & \wzero{} \\
 &  (flat)  & (flat,\omegam=0.3,\wa=0.5)  \\ 
\hline
 70 (real) GRBs (Amati+ 08) & 0.27$_{-0.18}^{+0.38}$  & $<$$-$0.3 (90\%)   \\
 156 (real) GRBs (Amati+ 13)  & 0.29$_{-0.15}^{+0.28}$  & $-$0.9$_{-1.5}^{+0.4}$  \\
 250 (156 real + 94 simulated) GRBs &  0.29$_{-0.12}^{+0.16}$ & $-$0.9$_{-1.1}^{+0.3}$   \\
 500 (156 real + 344 simulated) GRBs &  0.29$_{-0.09}^{+0.10}$  & $-$0.9$_{-0.8}^{+0.2}$ \\
\hline
 156 (real) GRBs, calibration &  0.30$_{-0.06}^{+0.06}$ &  $-$1.1$_{-0.30}^{+0.25}$ \\
 250 (156 real + 94 simulated) GRBs, calibration &  0.30$_{-0.05}^{+0.04}$ & $-$1.1$_{-0.20}^{+0.20}$  \\
 500 (156 real + 344 simulated) GRBs, calibration &  0.30$_{-0.03}^{+0.03}$ & $-$1.1$_{-0.15}^{+0.12}$ \\
\hline
\end{tabular} 
}}
\label{ta1}
\end{table}

 In the last 3 lines of Table~1 we report the estimates
of \omegam{} and \wzero{} derived from the present and expected future samples by assuming that
the \epeiso{} correlation is calibrated with a 10\% accuracy by using, e.g., the luminosity distances
provided by SNe--Ia, GRBs self-calibration  or the other methods shortly
described below.
The perspectives of
this method, combined with the expected increase of the number of GRBs in the sample, for the
investigation of the properties of ``dark energy," 
are shown in Fig.~7.

It is important to note that, as the number of GRBs in each $z$-bin increases, also the
feasibility and accuracy of the self-calibration of the \epeiso{}
correlation will improve.
Thus, the expected results shown in the last part of Table~1 and in 
Fig.~7 may be obtained even without the need of calibrating GRBs against other
cosmological probes.

\section{Further investigations and perspectives}

Table~1 shows a sharp increase of the accuracy attached to \omegam{} as a consequence 
of the increasing number of GRBs in the \epeiso{} plane. Currently, the main 
contribution to enlarge the GRB sample comes from joint detections by \swift{},
\fermi/GBM or Konus-WIND. Hopefully, these missions will continue to operate in
the next years, then providing us with an ``actual'' rate of $\sim$15--20 GRB/year. However, 
a real breakthrough in this field should come from next generation ($>$2017) 
missions capable of promptly pinpointing the GRB localization and of carrying out 
broad-band spectroscopy. We build our hopes, e.g., on the Japanese-led 
CALET/GBM experiment for the ISS\cite{Yamaoka09},
the Chinese-French mission SVOM 
\cite{Godet12}, the Lomonosov/UFFO \cite{Grossan12} experiment, an international project 
led by Korea and Taiwan, the WFM on-board LOFT, a mission currently under assessment study within the ESA/M3 programme\cite{Amati13c}. In Fig.~6 we show the confidence level contours in 
the \omegam--\omegal{} plane by using the real data as of 2012 and by adding to
them 113 and 363 simulated GRBs (resulting in two samples of 250 and 500 GRBs in total, 
respectively). The simulated data sets were obtained via Monte Carlo techniques by taking into account the slope,
normalization and dispersion of the observed \epeiso{} correlation, the
observed redshift distribution of GRBs and the distribution of the uncertainties
in the measured values of \epi{} and \eiso{}.
These simulations indicate that with a sample of 
250 GRBs (achievable in about 5 years from now) the accuracy in measuring 
$\Omega_M$ will be comparable to that currently provided by SNe data.

\begin{figure*}[t]
\centerline{\includegraphics[width=0.59\columnwidth,height=5.9cm]{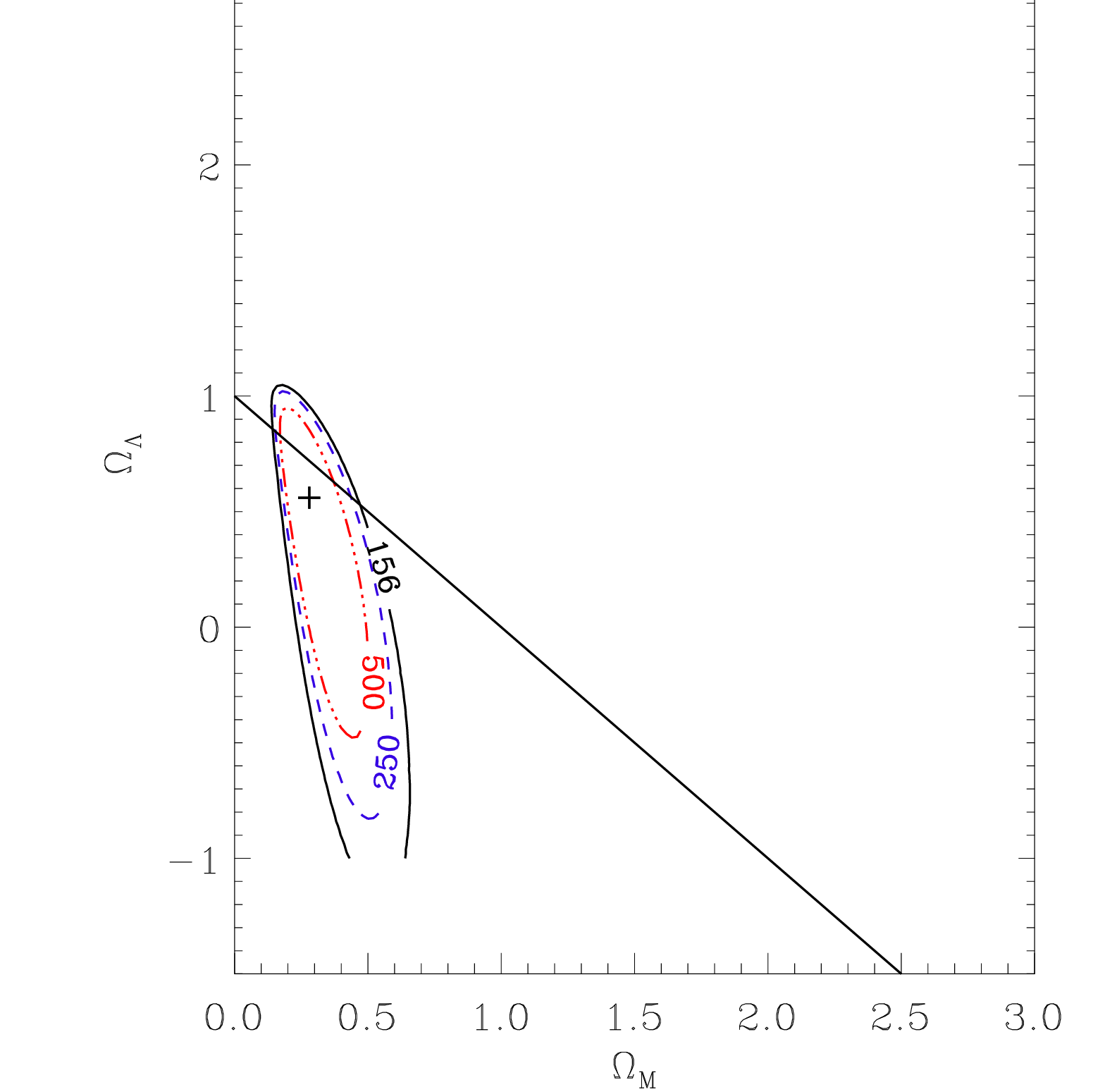}
\includegraphics[width=0.50\columnwidth,height=6.8cm]{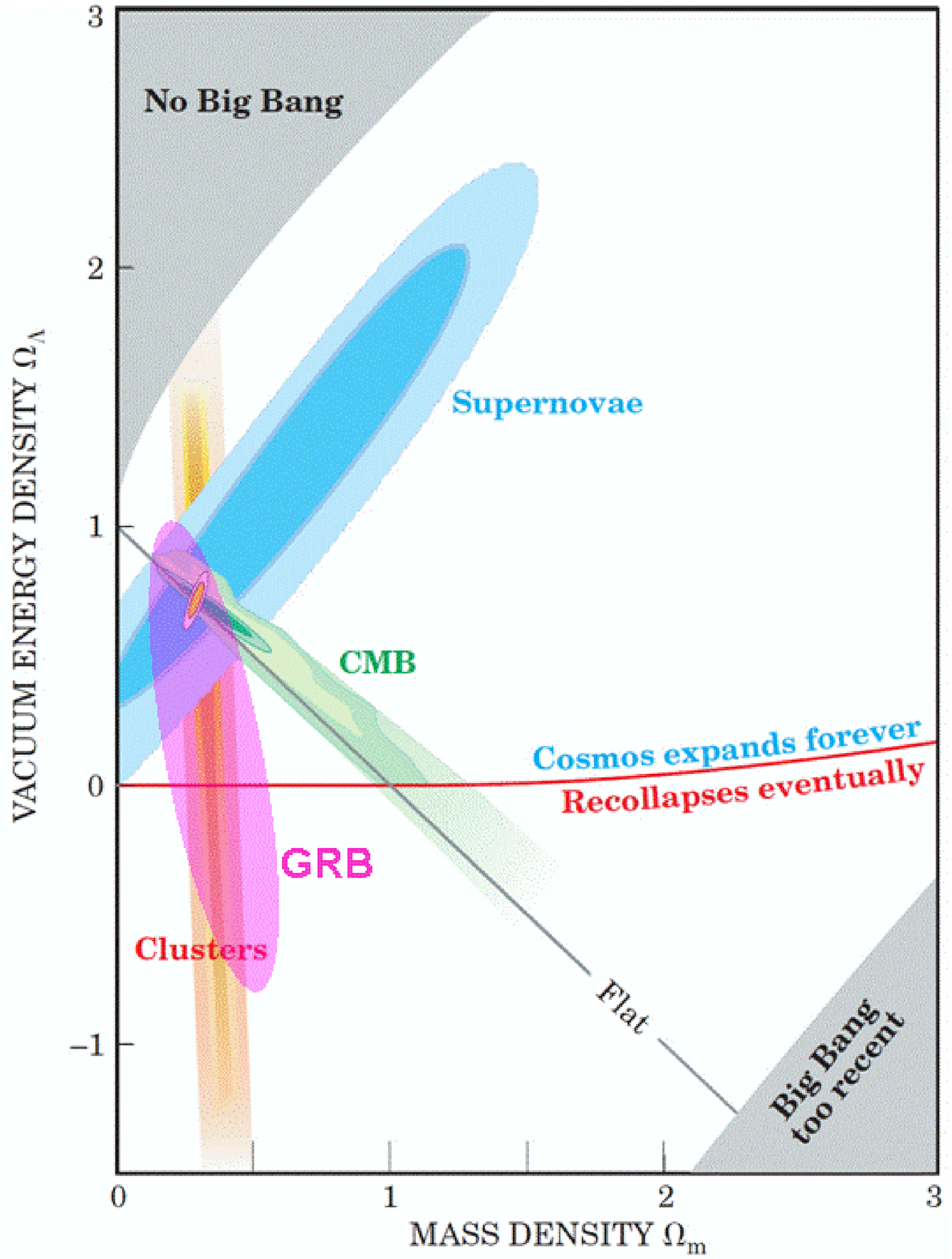}}
\caption{Left: 68\% confidence level contour in the \omegam{}--\omegal{} plane 
obtained by releasing the flat universe assumption with the sample of 156 GRBs 
available at the end of 2012 (black) compared to those expected in the next years 
with the increasing of GRBs in the sample (blue and red).\cite{Amati13b}
Right: 68\% confidence 
level contour in the \omegam{}--\omegal{} plane obtained by assuming a sample of 
250 GRBs expected in the near future (pink) compared to those from other 
cosmological probes (adapted from a figure by the Supernova Cosmology Project
\cite{Perlmutter03}).}
\end{figure*}

\begin{figure}[t]
\centerline{\includegraphics[width=0.55\columnwidth,height=6.0cm]{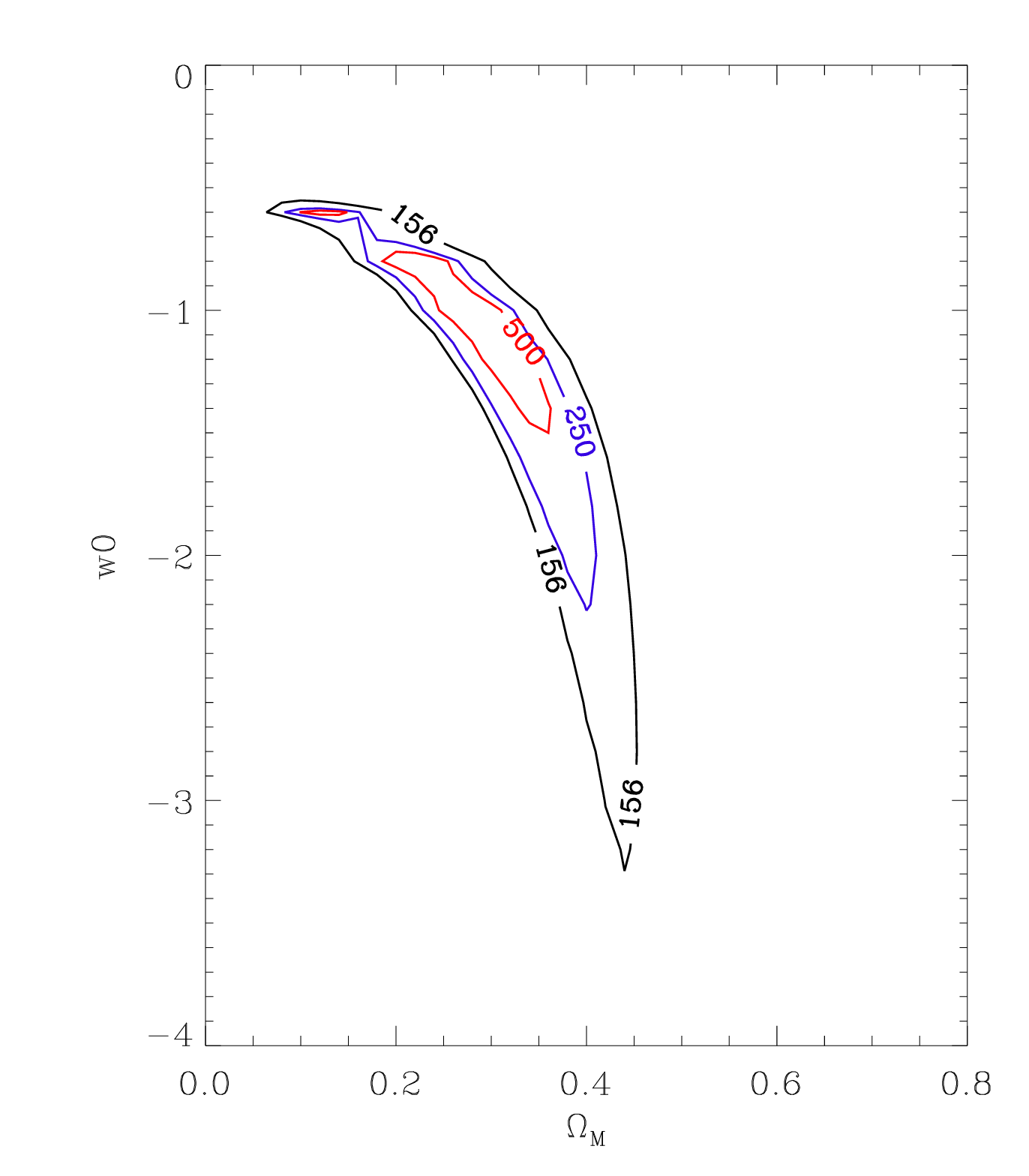}
\includegraphics[width=0.47\columnwidth]{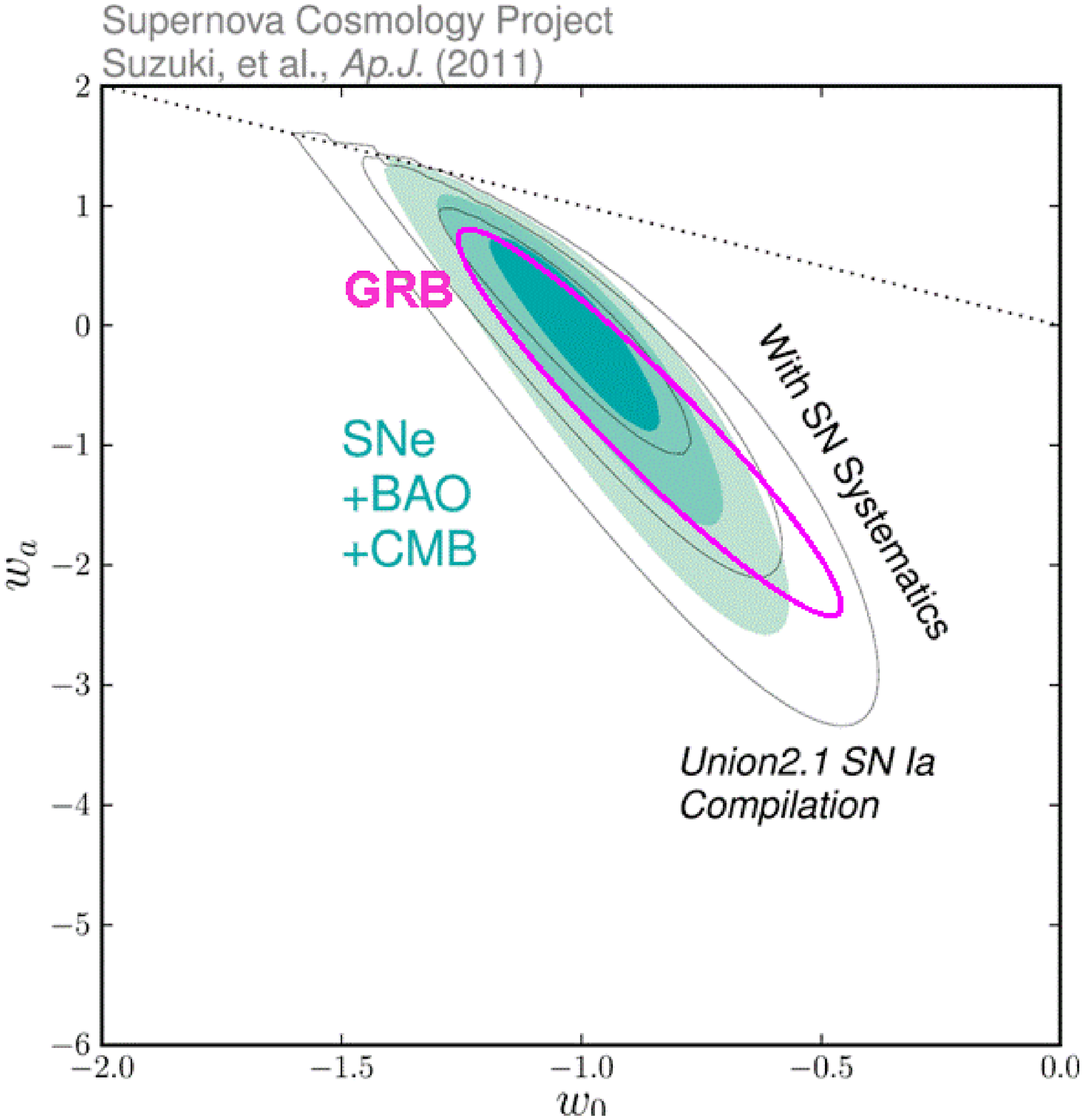}}
\caption{Left: 68\% confidence level contours in the \omegam{}--\wzero{} plane 
for a flat FLWR universe derived from the present and expected future samples by assuming that
the \epeiso{} correlation is calibrated with a 10\% accuracy by using, e.g., the luminosity distances
provided by SNe--Ia, self-calibration or the other methods described in the text.
Right: 68\% confidence 
level contour in the \wzero{}--\wa{} plane for a flat FLWR universe with
\omegam=0.3 obtained, by assuming that the \epeiso{} correlation is calibrated with a 10\% accuracy, using the
500 GRBs expected form next generation experiments (pink) compared to those from other 
cosmological probes (adapted from a figure by the Supernova Cosmology Project
\cite{Suzuki12}).}
\end{figure}

Several authors\cite{Kodama08,Capozziello10,Demianski12} are 
investigating the calibration of the \epeiso{} correlation at z $<$ 1.4 by using 
the luminosity distance versus redshift relation derived from SNe--Ia. The goal is 
to extend the SN--Ia Hubble diagram up to redshifts at which the luminosity distance 
is more sensitive to dark energy properties and evolution. The drawback of this 
approach is that, with this method, GRBs are no longer an independent cosmological 
probe. 

Other approaches include: cosmographic calibration of the \epeiso{} 
correlation\cite{Capozziello10,Demianski12}; 
self-calibration of the correlation with a large enough number of GRBs lying 
within a narrow ($\Delta$z $\sim$ 0.1-0.2) range of $z$ (promising, but requires a 
significant sample enlargement); combining the \epeiso{} correlation with 
other (less tight) GRB correlations \cite
{Schaefer07,Mosquera08,Cardone09}; extending the \epeiso{} correlation 
by involving other prompt or afterglow properties \cite{Dainotti11} 
aimed at reducing the dispersion of the
correlation (but with the risk of increasing systematics and lowering
the number of GRBs that can be used); testing alternative cosmologies vs.\ the standard
$\Lambda$CDM model\cite{Montiel11,Freitas11,Smale11,Lu11,Liang11,Diaferio11,
Demianski11,Capozziello12,Cardone12,Velten13,Wei13}. Particularly promising, are the 
perspectives
of the self-calibration method, which will become feasible and accurate as the number
of GRBs in the \epeiso{} plane will increase.\cite{Ghirlanda06}

Understanding the physical basis for the \epeiso{} correlation is also 
of fundamental importance for both GRB physics and cosmology, and several 
groups are working on this issue in the framework of the different scenarios that 
have been proposed for the origin of GRB prompt emission.
\cite{Zhang02} 

\section{Conclusions}

Due to their huge radiated energies GRBs can be observed up to $z\sim$ 10, 
therefore they are very powerful cosmological tools, complementary to other probes 
such as SN--Ia, clusters, or BAO. The correlation between spectral peak photon 
energy \epi\ and intensity (\eiso, \liso, \lpiso) is one of the most robust and 
intriguing properties of GRBs and a promising tool for measuring cosmological 
parameters. Analyses in the last years provide \underline{independent} 
evidence that, if we live in a flat universe, then \omegam{} is $<$ 1 at $>$99.9\% 
c.l.\ and around $\sim$0.3, consistent with current measurements 
obtained via different methodologies. The simultaneous operation of \swift, 
\fermi/GBM, Konus-WIND will increase the number of useful samples ($z$ + \epi) at a 
rate of 15--20 GRB/year, improving the accuracy in the estimate of cosmological 
parameters. Future GRB experiments (e.g., CALET/GBM, SVOM, Lomonosov/UFFO, LOFT/WFM) will increase dramatically the 
number of GRB usable in the \epeiso{} plane up to $z\sim 10$ (13.2 Gyrs in terms of 
look-back time) and therefore it will be possible to use them to follow the 
dependence on time (if any) of the density of vacuum energy since the early stages 
of the universe.

In conclusion, GRBs
have already provided a direct and independent measurement of \omegam{} and
simulations show that on a time scale of $\sim$5 years they will be able to
achieve a comparable accuracy to SNe--Ia. But the surplus value of GRBs is
that, in perspective, they can measure $w_0$ and
$w_a$, namely the evolution of dark energy with time. 

\section*{Acknowledgments}

The authors thank Brian Schmidt for his criticisms to the manuscript, which helped to 
improve it. The authors acknowledge support by the Italian Ministry for Education, 
University and Research through PRIN MIUR 2009 project on ``Gamma ray bursts: from 
progenitors to the physics of the prompt emission process" (Prot.\ 2009 ERC3HT).


\begin{thebibliography}{0}    
%
\bibitem{Perlmutter98}
S. Perlmutter, G. Aldering, M. Della Valle {\it et al.}, {\it Nature} {\bf 391} (1998) 51.
%
\bibitem{Perlmutter99}
S: Perlmutter, G. Aldering, G. Goldhaber {\it et al.}, {\it ApJ} {\bf 517} (1999) 565.
%
\bibitem{Riess98} 
A.G. Riess, V.A. Filippenko, P. Challis {\it et al.}, {\it AJ} {\bf 116} (1998) 1009.
%
\bibitem{Schmidt98} 
B.P. Schmidt, N.B. Suntzeff, M.M. Phillips {\it et al.}, {\it ApJ} {\bf 507} (1998) 46.
%
\bibitem{Tonry03} 
J. Tonry,  P.B. Schmidt, B. Barris {\it et al.}, {\it ApJ} {\bf 594} (2003) 1.
%
\bibitem{Knop03} 
R.A. Knop, G. Aldering, R. Amanullah {\it et al.}, {\it ApJ} {\bf 598} (2003) 102.
%
\bibitem{Astier06} 
P. Astier, J. Guy, N. Regnault {\it et al.}, {\it A\&A} {\bf 447} (2006) 31.
%
\bibitem{WW07} 
W.M. Wood-Vasey, G. Miknaitis, C.W. Stubbs {\it et al.}, {\it ApJ} {\bf 666} (2007) 694.
%
\bibitem{Kowalski08} 
M. Kowalski, D. Rubin, G. Aldering {\it et al.}, {\it ApJ} {\bf 686} (2008) 749.
%
\bibitem{Debernardis00} 
P. De Bernardis, P.A.R. Ade, J.J. Bock {\it et al.}, {\it Nature} {\bf 404} (2000) 955.
%
\bibitem{Spergel03} 
D.N. Spergel, L. Verde, H.V. Peiris {\it et al.}, {\it ApJS} {\bf 148} (2003) 175.
%
\bibitem{Dunkley09} 
J. Dunkley, E. Komatsu, M.R. Nolta {\it et al.}, {\it ApJS} {\bf 180} (2009) 306.
%
\bibitem{Komatsu11} 
E. Komatsu, K.M. Smith, J. Dunkley {\it et al.}, {\it ApJS} {\bf 192} (2011) 18.
%
\bibitem{Ade13}
P.A.R. Ade, N. Aghanim, C. Armitage-Caplan {\it et al.}, {\it A\&A}, submitted (arXiv:1303.5076).
%
\bibitem{Eisenstein05} 
D.J. Eisenstein, I. Zehavi, D.W. Hogg {\it et al.}, {\it ApJ} {\bf 633} (2005) 560.
%
\bibitem{Percival10} 
W.J. Percival, B.A. Reid, D.J. Eisenstein {\it et al.}, {\it MNRAS} {\bf 401} (2010) 2148.
%
\bibitem{Suzuki12} 
N. Suzuki, D. Rubin, C. Lidman {\it et al.}, {\it ApJ} {\bf 746} (2012) 85.
%
\bibitem{Chevalier01}
M. Chevallier and D. Polarski, {\it Int.\ J.\ Mod.\ Phys.} {\bf D10} (2001) 213.
%
\bibitem{Linder03} 
E.V. Linder, {\it Phys.\ Rev.\ Lett.} {\bf 90} (2003) 091301.
%
\bibitem{Riess04} 
A.G. Riess, L.G. Strolger, J. Tonry {\it et al.}, {\it ApJ} {\bf 607} (2004) 665.
%
\bibitem{Carroll06} 
S.M. Carroll, {\it Nature} {\bf 440} (2006) 7088.
%
\bibitem{Piran04} 
T. Piran, {\it Rev.\ Mod.\ Phys.} {\bf 76} (2004) 1143.
%
\bibitem{Meszaros06} 
P. M\'eszaros, {\it Rep.\ Prog.\ Phys.} {\bf 69} (2006) 2259.
%
\bibitem{Gehrels09}
N. Gehrels, E. Ramirez-Ruiz and D.B. Fox, {\it ARA\&A} {\bf 47} (2009) 567.
%
\bibitem{Zhang11}
B. Zhang, {\it Comptes Rendus Physique} {\bf 12} (2011) 206 (arXiv:1104.0932).
%
\bibitem{Salvaterra09} 
R. Salvaterra, M. Della Valle, S. Campana {\it et al.}, {\it Nature} {\bf 461} (2009) 1258.
%
\bibitem{Cucchiara11} 
A. Cucchiara, S.B. Cenko, J.S. Bloom {\it et al.}, {\it ApJ} {\bf 743} (2011) 154.
%
\bibitem{Riess07} 
A.G. Riess, L.-G. Strolger, S. Casertano {\it et al.}, {\it ApJ} {\bf 659} (2007) 98.
%
\bibitem{Jones13}
D.O. Jones, S.A. Rodney, A.G. Riess {\it et al.}, {\it ApJ} {\bf 768} (2013) 166.
%
\bibitem{Dellavalle92}
M. Della Valle and N. Panagia, {\it ApJ} {\bf 104} (1992) 696.
%
\bibitem{Folatelli10}
G. Folatelli, M.M. Phillips, C.R. Burns {\it et al.}, {\it ApJ} {\bf 139} (2010) 120.
%
\bibitem{Amati08}
L. Amati, C. Guidorzi, F. Frontera {\it et al.}, {\it MNRAS} {\bf 391} (2008) 577.
%
\bibitem{Dellavalle09}
M. Della Valle and L. Amati, {\it AIP Conf.\ Proc.} {\bf 1053} (2009) 299.
%
\bibitem{Amati02}
L. Amati, F. Frontera, M. Tavani {\it et al.}, {\it A\&A} {\bf 390} (2002) 81.
%
\bibitem{Amati06}
L. Amati, {\it MNRAS} {\bf 372} (2006) 233.
%
\bibitem{Band93}
D. Band, J. Matteson, L. Ford {\it et al.}, {\it ApJ} {\bf 413} (1993) 281.
%
\bibitem{Guetta07}
D. Guetta and M. Della Valle, {\it ApJ} {\bf 657} (2007) L73.
%
\bibitem{Amati07}
L. Amati, M. Della Valle, F. Frontera {\it et al.}, {\it A\&A} {\bf 463} (2007) 913.
%
\bibitem{Ghirlanda13}
G. Ghirlanda, G. Ghisellini, R. Salvaterra {\it et al.}, {\it MNRAS} {\bf 428} (2013) 1410.
%
\bibitem{Amati13a} 
Amati, L. \& Della Valle, M., {\it Astron.\ Rev.} {\bf 8}, (2013) 90.
%
\bibitem{Amati13b}
L. Amati, M. Della Valle, F. Frontera {\it et al.}, in prep.
%
\bibitem{Zhang02}
B.M. Zhang and P. M\'esz\'aros,
{\it ApJ} {\bf 581} (2002) 1236.
%
\bibitem{Lamb04}
D.Q. Lamb, G.R. Ricker, J.-L. Atteia {\it et al.},
{\it NewAR} {\bf 48} (2004) 423.
%
\bibitem{Soderberg06} 
A.M. Soderberg, S.R. Kulkarni, E. Nakar, E. {\it et al.}, {\it Nature} {\bf 442} (2006) 1014.
%
\bibitem{Band05}
D. Band and R.D. Preece, {\it ApJ} {\bf 627} (2005) 319.
%
\bibitem{Nakar05}
E. Nakar and T. Piran, {\it MNRAS} {\bf 360} (2005) L73.
%
\bibitem{Ghirlanda05}
G. Ghirlanda, G. Ghisellini and C. Firmani, {\it MNRAS} {\bf 361} (2005) L10.
%
\bibitem{Bosnjak08}
Z. Bosnjak, A. Celotti, F. Longo, G. Barbiellini, {\it MNRAS} {\bf 384} (2008) 599.
%
\bibitem{Ghirlanda08}
G. Ghirlanda, L. Nava, G. Ghisellini, C. Firmani, J.I. Cabrera, {\it MNRAS} {\bf 387} (2008) 319.
%
\bibitem{Nava11}
L. Nava, G. Ghirlanda, G. Ghisellini, A. Celotti, {\it A\&A} {\bf 530} (2011) 21.
%
\bibitem{Amati09} 
L. Amati, F. Frontera and C. Guidorzi, {\it A\&A} {\bf 508} (2009) 173. 
%
\bibitem{Ghirlanda10}
G. Ghirlanda, L. Nava and G. Ghisellini, {\it A\&A} {\bf 511} (2010) 43.
%
\bibitem{Sakamoto11}
T. Sakamoto, S.D. Barthelmy, W.H. Baumgartner {\it et al.}, {\it ApJS} {\bf 195} (2011) 2.
%
\bibitem{Lu12}
R. Lu, J. Wei, E. Liang {\it et al.},
{\it ApJ} {\bf 756} (2012) 112.
%
\bibitem{Frontera12}
F. Frontera, L. Amati, J.J.M. in't Zand {\it et al.}, 
{\it ApJ} {\bf 754} (2012) 138.
%
\bibitem{Basak13}
R. Basak and A.R. Rao, {\it MNRAS} in press.
%
\bibitem{Ghirlanda04}
G. Ghirlanda, G. Ghisellini and D. Lazzati, 
{\it ApJ} {\bf 616} (2004) 331. 
%
\bibitem{Dai04} 
Z.G. Dai, E.W. Liang and D. Xu, {\it ApJ} {\bf 612} (2004) L101. 
%
\bibitem{Sari99} 
R. Sari, T. Piran and J.P. Halpern, {\it ApJ} {\bf 519} (2009) L17.
%
\bibitem{Ghirlanda06}
G. Ghirlanda, G. Ghisellini and C. Firmani,
{\it New J. Phys.} {\bf 8} (2006) 123. 
%
\bibitem{Schaefer07} 
B.E. Schaefer, {\it ApJ} {\bf 660} (2007) 16. 
%
\bibitem{Kodama08} 
Y. Kodama, D. Yonetoku, T. Murakami {\it et al.}, {\it MNRAS} {\bf 391} (2008) L1. 
%
\bibitem{Demianski12} 
M. Demianski, E. Piedipalumbo, C. Rubano, P. Scodellaro, {\it MNRAS} {\bf 426} (2012) 1396. 
%
\bibitem{Panaitescu06} 
A. Panaitescu, P. M\'esz\'aros, D.N. Burrows {\it et al.}, {\it MNRAS} {\bf 369} (2006) 2059.
%
\bibitem{Curran08} 
P.A. Curran, A.J. van der Horst and R.A.M.J. Wijers, {\it MNRAS} {\bf 386} (2008) 859.
%
\bibitem{Liang08} 
E.-W. Liang, J.L. Racusin, B. Zhang, B. {\it et al.}, {\it ApJ} {\bf 675} (2008) 528.
%
\bibitem{Dagostini05} 
G. D'Agostini, {\it ArXiv preprint} (2005) physics/0511182.
%
\bibitem{Reichart01}
D.E. Reichart, {\it ApJ} {\bf 553} (2001) 235.
%
\bibitem{Yamaoka09}
K. Yamaoka, A. Yoshida, T. Kotani {\it et al.}, {\it AIP Conf.\ Proc.} {\bf 1133} (2009) 88.
%
\bibitem{Godet12}
O. Godet, J. Paul, J.Y. Wei {\it et al.}, {\it SPIE Conf.\ Proc.} {\bf 8443} (2012) 10.
%
\bibitem{Grossan12}
B. Grossan, I.H. Park, S. Ahmad {\it et al.}, {\it SPIE Conf.\ Proc.} {\bf 8443} (2012) 2.
%
\bibitem{Amati13c}
L. Amati, E. Del Monte, V. D'Elia {\it et al.}, {\it Nucl.\ Phys.B - Proc.\ Suppl.} {\bf 239} (2013) 109.
%
\bibitem{Perlmutter03}
S. Perlmutter, {\it Physics Today} {\bf 56} (2003) 53.
%
\bibitem{Capozziello10} 
S. Capozziello and L. Izzo, {\it A\&A} {\bf 519} (2010) A73. 
%
\bibitem{Mosquera08} 
H.J. Mosquera Cuesta, M.H. Dumet M.H. and C. Furlanetto, {\it JCAP} {\bf 7} (2008) 4.
%
\bibitem{Cardone09} 
V.F. Cardone, S. Capozziello and M.G. Dainotti, {\it MNRAS} {\bf 400} (2009) 775.
%
\bibitem{Dainotti11} 
M.G. Dainotti, M. Ostrowski and R. Willingale, {\it MNRAS} {\bf 418} (2011) 2202.
%
\bibitem{Montiel11}
A. Montiel and N. Bret\'on, {\it JCAP} {\bf 8} (2011) 23.
%
\bibitem{Freitas11}
R.C. Freitas, S.V.B. Goncalves and H.E.S. Velten,
{\it Physics Letters B} {\bf 3} (2011) 209.
%
\bibitem{Smale11}
P.R. Smale, {\it MNRAS} {\bf 418} (2011) 2779.
%
\bibitem{Lu11}
J. Lu, Y. Wang, Y. Wu and T. Wang, {\it EPJC} {\bf 71} (2011) 1800.
%
\bibitem{Liang11}
N. Liang, P.-X. Wu and Z.-H. Zhu, {\it RAA} {\bf 11} (2011) 9.
%
\bibitem{Diaferio11}
A. Diaferio, L. Ostorero and V. Cardone, {\it JCAP} {\bf 10} (2011) 8.
%
\bibitem{Demianski11} 
M. Demianski, E. Piedipalumbo and C. Rubano, {\it MNRAS} {\bf 411} (2011) 1213. 
%
\bibitem{Capozziello12} 
S. Capozziello, L. Consiglio, M. De Laurentis {\it et al.}, {\it ArXiv preprint} (2012) 1206.6700.
%
\bibitem{Cardone12}
V.F. Cardone, S. Camera and A.Diaferio, {\it JCAP} {\bf 2} (2012) 30.
%
\bibitem{Velten13}
H. Velten, A. Montiel and S. Carneiro, {\it MNRAS} {\bf 431} (2013) 3301.
%
\bibitem{Wei13}
J.-J. Wei, X.-F. Wu and F. Melia, {\it ApJ} {\bf 772} (2013) 43.
%
\end{thebibliography}
\end{document}